\documentclass{aa}
\usepackage{psfig}
\begin{document}

\thesaurus{01
 (09.13.2;  
            11.03.2;  
            11.06.1;  
            11.09.4;  
            11.19.3   
            13.19.1)}  

   \title{A sensitive 
search for CO emission from faint blue galaxies at $z\sim 0.5$}
 
   \author{C. D. Wilson
          \inst{1}
\and F. Combes
          \inst{2}
          }
 
   \offprints{C. D. Wilson}
 
   \institute{Department of Physics \& Astronomy, McMaster University,
              Hamilton, Ontario L8S 4M1, Canada; wilson@physics.mcmaster.ca
\and DEMIRM, Observatoire de Paris, 61 Av. de l'Observatoire, F-75014 
Paris, France; bottaro@mesioa.obspm.fr  }
 
   \date{Received ; accepted }

\titlerunning{Search for CO emission from faint blue galaxies}
\authorrunning{Wilson \& Combes}

   \maketitle

   \begin{abstract}

We have obtained sensitive upper limits on the CO J=2-1 and CO J=3-2 
emission lines for five faint blue galaxies with redshifts $z\sim 0.5$
using the IRAM 30~m telescope.
These observations would have been able to detect the
luminous infrared galaxy IRAS F10214+4724 if it were located at this
redshift and unlensed. However, they are not sensitive enough to detect the
prototype starburst galaxy M82 or the HII galaxy UM448 if they were
located at this redshift. Our upper limits for the CO emission are
consistent with between 19\% and 66\% of the total galactic mass being
in the form of molecular hydrogen, and thus shed little light on the
ultimate fate of these galaxies.

\keywords{interstellar medium: molecules -- 
          galaxies: compact --
          galaxies: formation --
          galaxies:  ISM --
          galaxies: starburst -- 
          radio lines: galaxies }

   \end{abstract}

%

\section{Introduction}

A large population of faint blue galaxies 
at moderate redshifts was first identified by Tyson (\cite{tyson}).
These galaxies are puzzling both because of their large
space density (30 times that of ordinary bright galaxies,
Lilly et al. \cite{lilly}; Babul \& Rees \cite{babul}) and
their lack of obvious bright counterparts in the local universe.
Recent studies have provided us with a detailed look at a sample
of these galaxies at $z=0.1-0.6$ (Koo et al. \cite{koo94}, \cite{koo95};
Guzman et al. \cite{guzman}). The galaxies have star formation rates of
1-20 M$_\odot$ yr$^{-1}$ (Koo et al. \cite{koo95}),
comparable to the total star formation rates of present
day disk galaxies (Kennicutt \cite{kennicutt}). 
However, they are very
compact, with half-maximum diameters of only 2-4 kpc (Koo et al. \cite{koo94}).
Although their blue luminosities are also comparable to local field galaxies
($M_B \sim -21$), their total masses are estimated at only $1-5\times 10^9$
M$_\odot$, much smaller than the typical masses of normal elliptical
or spiral galaxies ($\sim 10^{11}$ M$_\odot$) (Guzman et al. \cite{guzman}).
These observations suggest the galaxies are  dwarf galaxies
near their peak luminosity after undergoing a major burst of star formation.
Since star formation is intricately linked with the presence of molecular gas 
in the local universe, these galaxies are likely to contain
significant amounts of molecular gas to provide the fuel for the
observed starbursts.

The detection of CO at $z=2.28$ in the luminous infrared galaxy
IRAS F10214+4724 (Brown \& Vanden Bout \cite{brown}; Solomon et al. 
\cite{sol92}) has 
stimulated numerous searches for CO emission from galaxies at moderate 
to high redshifts (Wiklind \& Combes \cite{wik94b}; Evans et al. \cite{evans}).
However, these searches were less successful than expected, and only highly 
amplified objects have been detected: the Cloverleaf quasar at $z=2.56$
(Barvainis et al. \cite{barvainis}) and BR1202-0725 at $z=4.69$ (Ohta et al. 
\cite{ohta}; Omont et 
al. \cite{omont}). 
Recently, 
Scoville et al. (\cite{scoville}) reported the detection of the first 
non-lensed object at $z=2.394$, the weak radio galaxy 53W002.
The derived molecular mass is so high 
($7.4 \times 10^{10}$ M$_\odot$ with a standard CO-to-H$_{2}$ 
conversion factor, and even more 
if the metallicity is low) that it 
constitutes between 30 and 80\% of the total dynamical mass,
depending on the unknown inclination.
The lensing that can amplify CO emission to a level that is detectable
with current instruments can also be produced 
by galaxy clusters. A search for CO emission from four giant arcs
in clusters has been reported by Casoli et al. (\cite{casoli}), 
and they detected
one of them at $z= 0.725$. Another sensitive technique 
for probing the cold molecular interstellar medium 
at high redshifts is via absorption lines
(e.g. Combes \& Wiklind \cite{combes}). 
Four systems have been detected in more than a dozen molecules in absorption
with $z$ between 0.25 and 0.9 
(Wiklind \& Combes \cite{wik94a}, \cite{wik95}, \cite{wik96a}, \cite{wik96b}).

In this letter, we present the results of
a survey for CO emission from five 
faint blue galaxies at redshifts $z\sim 0.5$ selected from
the sample of Koo et al. (\cite{koo95}). This survey differs from
previous surveys in targeting galaxies with known
star formation rates, as opposed to damped Lyman $\alpha$
systems (Wiklind \& Combes \cite{wik94b}) or distant radio galaxies
(Evans et al. \cite{evans}). Unfortunately, as in previous surveys,
we have achieved only upper limits to the CO
emission from our target objects.

\section{Observations and data reduction}

\subsection{Observations}

   \begin{figure}
%
%
\psfig{figure=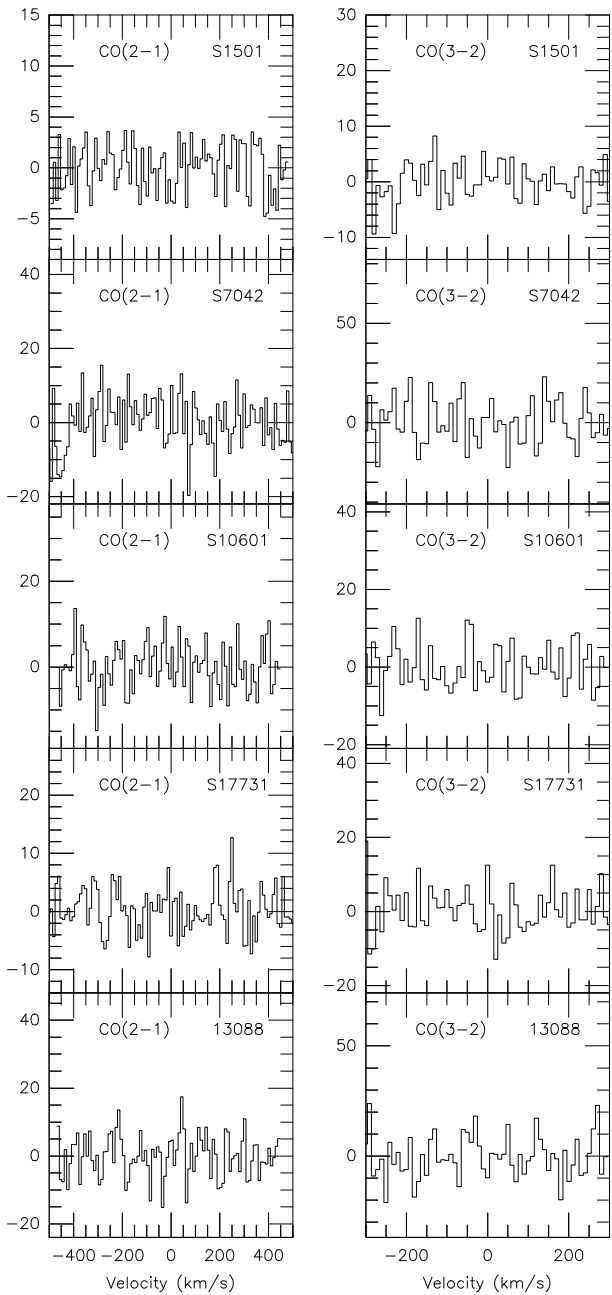,width=8.5cm,bbllx=39mm,bblly=43mm,bburx=137mm,bbury=244mm}
     \caption[]{$^{12}$CO J=2-1 and J=3-2 spectra for the five
faint blue galaxies. The observed CO transition is marked in each
spectrum. The temperature scale is T$_{mb}$ in mK and
the velocity resolution is 10 km s$^{-1}$. No galaxies
were detected.}\label{fig-1}
    \end{figure}

Of the 17 galaxies in the Koo et al. (\cite{koo95}) 
sample of faint blue galaxies,
nine have redshifts such that the CO J=2-1 and J=3-2 lines
are accessible with the IRAM receivers. Three of these nine galaxies
are somewhat more extended than the other six and indeed
appear non-stellar in ground-based optical observations (Koo et al.
\cite{koo95}). We excluded these three galaxies from our sample 
on the grounds that they may be somewhat more massive objects,
akin perhaps to small spiral galaxies. 
Observations of five of the remaining six galaxies 
were obtained with the IRAM 30 m telescope in two separate
observing runs in January 1996 and June 1997.  The sixth galaxy,
SA57-5482, was not observed due to time constraints.
The half-power beam width is 17$^{\prime\prime}$
at 2--mm and 12$^{\prime\prime}$ at 1.3--mm. We used the 2-- and 1.3--mm  SiS
receivers to observe both lines simultaneously. The receivers were
all used in single sideband mode, and the typical system temperatures
were 250-500 K for the 2--mm receiver and 350-700 K for the 1.3--mm receiver,
in T$_A^*$ scale (or, on average, 600 K 
and 1200 K in T$_{MB}$ scale, respectively).
The backends were essentially two 1MHz-filter-banks, of 512 channels each,
and in addition an auto-correlator; the spectra have been smoothed to
10 km s$^{-1}$ resolution. The observations
were made using a nutating secondary with a beam throw of 1.5$^\prime$.
The pointing was checked every two hours and the pointing accuracy was
estimated to be 3$^{\prime\prime}$ rms.

\subsection{Data reduction and analysis}

   \begin{table*}
      \caption[]{CO upper limits for faint blue galaxies}
\label{tbl-1}
\begin{flushleft}
\begin{tabular}{llllllllll}
\noalign{\smallskip}
\hline
\noalign{\smallskip}
Galaxy & RA(1950) & DEC(1950) & z & $\Delta V$ & 
time & $\sigma(2-1)$ & $\sigma(3-2)$ 
& $I_{CO}(3-2)$ & $L_{CO}(3-2)$ \\
& ($h$ $m$ $s$) & ($^o$ $^\prime$ $^{\prime\prime}$) & & (km s$^{-1}$) 
& (hr) & (mK) & (mK) &
(K km s$^{-1}$) & (K km s$^{-1}$ pc$^2$) \\
\noalign{\smallskip}
\hline
\noalign{\smallskip}
SA57-1501 &13:05:43.9 &29:24:59 &0.4993 & 143 & 9.3 & 2.2 & 3.7 & 0.42 & 2.6$\times 10^9$ \\
SA57-7042 &13:05:03.2 &29:34:26 &0.5250 & 273 & 2.4 & 6.3 & 10.8 & 1.69 & 1.1$\times 10^{10}$\\
SA57-10601 &13:06:25.0 &29:39:39 &0.4384 & 101 & 1.8 & 5.3 & 5.5 & 0.52 & 2.8$\times 10^9$\\
SA57-17731 &13:06:15.5 &29:50:47 &0.6612 & 242 & 3.9 & 3.7 & 5.9 & 0.87 & 7.4$\times 10^9$ \\
Herc-1-13088 &17:19:04.4 &50:04:00 &0.4357 & 106 & 4.3 & 6.8 & 9.7 & 0.95 & 5.1$\times 10^9$
\\
\noalign{\smallskip}
\hline
\end{tabular}
\end{flushleft}
\begin{list}{}{}
\item[] Temperatures are given in the main beam scale
\end{list}
   \end{table*}

The spectra were first inspected, and any spectrum showing baseline
curvature or other artifacts was discarded. 
The remaining spectra were averaged together, weighted by their rms noise.
A first order baseline was removed from each average spectrum, and
the spectra were smoothed 
to a resolution of 10 km s$^{-1}$ to produce
the final spectra (Fig.~\ref{fig-1}). 
The final temperatures (and rms noise in Table 1)
have been converted to the $T_{MB}$ temperature scale ($\eta_{MB} = 0.45$
at 230 GHz, 0.59 at 150 GHz).

Upper limits to the integrated CO intensity were derived using the
rms noise measured from the CO spectra and the velocity widths
obtained from measurements of optical emission lines (Koo et al. \cite{koo95}).
We adopt as the 3$\sigma$ upper limit to the CO intensity
$$I_{CO} \le {{3\sigma \Delta V} \over {\sqrt{N_{chan}}}}
\hskip6pt \rm{K\hskip3pt km \hskip3pt s^{-1}}$$
(Wiklind \& Combes \cite{wik94b}), 
where $\sigma$ is the rms noise in K measured in our 10 km s$^{-1}$
channels, $\Delta V$ is the velocity width of the CO line, here
taken to be the full-width half-maximum of the optical lines,
and $N_{chan} = \Delta V/10$ km s$^{-1}$ is the number of channels in the 
velocity width.
The CO luminosity for a source at high redshift is given by
$$L_{CO} = 23.5 I_{CO} \Omega_B {{D_L^2}\over {(1+z)^3}} \hskip6pt
\rm{K\hskip3pt km \hskip3pt s^{-1}\hskip3pt pc^2}$$
where $\Omega_B$ is the area of the main beam in square arcseconds
and $D_L = (c/H_o q_o^2)[q_o z + (q_o-1)(\sqrt{1+2q_oz}-1)]$ 
is the luminosity distance in Mpc (Wiklind \& Combes \cite{wik94b}). We adopt
$q_o=0.5$ and $H_o = 70$ km s$^{-1}$ Mpc$^{-1}$ in this paper.
Table~\ref{tbl-1} gives the position, redshift, and velocity
width obtained from the optical emission lines (Koo et al. \cite{koo95}), as
well as 
the integration time, the rms noise for each line, and the CO
integrated intensity and CO luminosity calculated from the CO J=3-2 
upper limit.

\section{Discussion}

Our upper limits to the CO flux are comparable to the best
upper limits  in the literature for moderate to high redshift 
objects. For example,
the detections of CO J=3-2 emission at high redshift are
6.7 Jy km s$^{-1}$ for IRAS F10214+4724 (Radford et al. \cite{radford}) 
and 8.1 Jy km s$^{-1}$ for the Cloverleaf (Barvainis et al. \cite{barvainis}),
while our 3$\sigma$ upper limits range from 2 to 8
Jy km s$^{-1}$. If our galaxies had comparable
CO fluxes to IRAS F10214+4724 or the Cloverleaf quasar, we would
have detected them with our observations. In addition, if we assume
the amplification due to lensing is a factor of 10 in the two 
high redshift galaxies, their CO luminosities $L_{CO}$ (converted to our
cosmology) are $8.9\times 10^9$ and $1.3\times 10^{10}$ K km s$^{-1}$ pc$^2$,
respectively. Thus, we would have detected either of these two
galaxies, unlensed, at a redshift of $z \sim 0.5$.

Since these faint blue galaxies are thought to be distant counterparts to
HII galaxies, we should also compare our upper limits with CO
observations of nearby dwarf galaxies. The CO J=1-0 luminosities
of the starburst galaxy M82 and the HII galaxy UM448 are both
$L_{CO} \sim 5\times 10^8$ K km s$^{-1}$ pc$^2$ (calculated
from Young et al. \cite{young}; Sage et al. \cite{sage}). 
Unfortunately, our best
upper limits are still a factor of 4-8 larger than the luminosities
of these nearby dwarf galaxies, and so we would not have detected M82
or UM448 at $z\sim 0.5$. 

For galaxies in the local universe with near-solar metallicities and
normal rates of star formation (i.e. not starburst galaxies), the
mass of molecular hydrogen gas is related to the CO luminosity in
the J=1-0 line by $M_{H_2} = 4.8 L_{CO}$ M$_\odot$ (i.e.
Solomon et al. \cite{sol87}). Since we have observed the CO J=2-1 and
J=3-2 lines, we must consider the excitation of the gas in
estimating molecular gas masses. In galactic nuclei, the three transitions
have similar strengths (Braine \& Combes \cite{braine}; G\"usten et al. 
\cite{gusten}),
while in the disks of spiral galaxies, the J=2-1/1-0 ratio is typically
0.5-0.7 (Braine \& Combes \cite{braine}; Sakamoto et al. \cite{sakamoto}) 
and the
3-2/2-1 ratio is 0.7 (Wilson et al. \cite{wilson97}). Given the compact
and starburst nature of these systems, we will assume that the three
lowest CO line ratios have equal strengths, in which case our CO J=3-2
observations give the best upper limit to the molecular mass. If in fact
these galaxies are more like normal spirals in their excitation, our
gas masses will be underestimated by a factor of about two. 

The CO-to-H$_2$ conversion factor is known
to depend on the metallicity of the galaxy (Wilson \cite{wilson95}) and is
also expected to be different in the unusual conditions present in
a starburst galaxy. The metallicities of these galaxies are estimated
to be $\sim 0.7$ solar (Guzman et al. \cite{guzman}), which implies a roughly
normal CO-to-H$_2$ conversion factor (Wilson \cite{wilson95}). 
Assessing the impact
of the starburst environment is more difficult, although Solomon et al. 
(\cite{sol97})
suggest that the conversion factor is roughly four times smaller
in luminous infrared galaxies than in normal spirals. To take into
account the intense star formation in these galaxies, we will
calculate the H$_2$ mass using $M_{H_2} = 1.2 L_{CO}$ M$_\odot$; note that
this may underestimate the gas masses by up to a factor of four if
the star formation environment is relatively normal.

   \begin{table}
      \caption[]{Comparison of virial and gas masses}
\label{tbl-2}
\begin{flushleft}
\begin{tabular}{lll}
\noalign{\smallskip}
\hline
\noalign{\smallskip}
Galaxy & $M_{vir}$ & $M_{H_2}$ \\
& $10^9$ M$_\odot$ & $10^9$ M$_\odot$ \\
\noalign{\smallskip}
\hline
\noalign{\smallskip}
SA57-1501 & 17 & $<$3.2 \\
SA57-7042 & 60 & $<$14 \\
SA57-10601 & 8.4 & $<$3.4 \\
SA57-17731 & 48 & $<$8.9 \\
Herc-1-13088 & 9.2 & $<$6.1 \\
\noalign{\smallskip}
\hline
\end{tabular}
\end{flushleft}
   \end{table}

Table~\ref{tbl-2} compares the upper limits to the gas masses with the virial
masses of the galaxies. The virial masses were calculated as in
Guzman et al. (\cite{guzman}), and then scaled up by a factor
of 4 to account for possible underestimate of the total mass of
the galaxy using the optical data (see discussion in Guzman et al. 
\cite{guzman}). (These virial mass estimates will be even more uncertain
if these objects are predominantly disk-like rather than spheroidal
systems, as assumed in the calculation.)
Since the radii of our galaxies have not
been measured, we adopt the average effective radius $R_e = 0.96$ kpc
(converted to our cosmology) measured for seven faint blue galaxies by
Koo et al. (\cite{koo94}). Table~\ref{tbl-2} 
shows that, even with our conservative assumptions,
which increase the total mass of the galaxy and decrease the gas mass,
our upper limits are consistent with the faint blue galaxies containing
between 19\% and 66\% of their total mass in the form of molecular gas.
Depending on how much gas  they can
retain  to fuel later star formation episodes, these galaxies
may either be the
progenitors of present day dwarf elliptical galaxies (if most
of the gas is lost in the current burst, Koo et al. \cite{koo95})
 or HII galaxies (if a significant fraction of their gas survives
the current burst).
Unfortunately, these observations provide no constraint on the interesting
question of whether or not these galaxies still contain large quantities
of molecular gas.  

\section{Conclusions}

We have obtained sensitive upper limits on the CO J=2-1 and CO J=3-2 
emission lines for five faint blue galaxies with redshifts $z\sim 0.5$
for which accurate redshifts and linewidths had been determined
previously from optical observations. Our upper limits
are comparable to the best upper limits published for
moderate to high redshift objects in the literature, and are sensitive
enough to have detected the
luminous infrared galaxy IRAS F10214+4724 if it were located at this
redshift and unlensed. However, our sensitivity is insufficient
to detect either the 
prototype starburst galaxy M82 or the HII galaxy UM448 if they were
located at this redshift. Our upper limits for the CO emission are
consistent with these galaxies containing between 19\% and 66\% of their
 total mass in the form of molecular hydrogen. Since whether
these galaxies will evolve into dwarf ellipticals or retain enough
gas to burst again as HII galaxies depends on their ability to
retain gas during the observed burst of star formation, our observations
are unable to provide any useful information on the 
ultimate fate of these galaxies.

\begin{acknowledgements}
The research of C.D.W. is supported by a grant from NSERC (Canada).
\end{acknowledgements}

\end{document}